\documentclass[12pt]{article}
\usepackage{newtxtext,newtxmath}
\usepackage{graphicx}
\usepackage[letterpaper,margin=1in]{geometry}
\renewenvironment{abstract}
	{\quotation}
	{\endquotation}

\date{}

\makeatletter
\renewcommand{\fnum@figure}{\textbf{Figure \thefigure}}
\renewcommand{\fnum@table}{\textbf{Table \thetable}}
\makeatother
\usepackage{scicite}
\usepackage{url}

\def\scititle{
	Nanomechanical detection\\of vortices in an electron fluid
}

\title{\bfseries \boldmath \scititle}

\author{
	Andrey~A.~Shevyrin$^{1\ast}$,
	Askhat~K.~Bakarov$^{1}$,
	Arthur~G.~Pogosov$^{1,2}$\and
	\small$^{1}$Rzhanov Institute of Semiconductor Physics, Novosibirsk, 630090, Russia.\and
	\small$^{2}$Department of Physics, Novosibirsk State University, Novosibirk, 630090, Russia.\and
	\small$^\ast$Corresponding author. Email: shevandrey@isp.nsc.ru
}

\begin{document}

\maketitle
\begin{abstract} \bfseries \boldmath

Electron vortices are the quintessential signature of a viscous electron fluid. For decades, their detection relied on indirect transport measurements with persistently debated interpretations. Recently, scanning magnetometry enabled direct visualization, yet these techniques demand considerable sophistication. Here we introduce a conceptually different and inherently simpler paradigm based on nanomechanics. By integrating a circular cavity into a suspended resonator, we create a vortex whose circulating current generates a magnetic moment. In an in-plane magnetic field, this moment experiences a torque, driving vibrations that directly reveal the vortex’s presence and nature. We detect ballistic and hydrodynamic vortices and trace their temperature-driven crossover. Our work establishes nanomechanics as a platform for electron hydrodynamics, showing that viscosity—subtle in transport—is one of the dominant factors shaping nanoelectromechanical response.
\end{abstract}

\noindent
The idea of electrons flowing as a viscous fluid — highlighted by the pioneering work of Gurzhi \cite{Gurzhi1968} — is nearly as old as Fermi liquid theory itself. Intuitively appealing, it offers a simple description of complex collective electron motion \cite{Varnavides2023,Fritz2024,Narozhny2022,Polini2020} — one where electrons scatter off each other more frequently than off boundaries, impurities, or phonons. The hydrodynamic effects expected in this framework include resistance drop with temperature increase (Gurzhi effect), Poiseuille flow \cite{Sulpizio2019,Ku2020,Vool2021,Gusev2020,KrishnaKumar2017}, and, most intuitively, vortices \cite{Bandurin2016,Gupta2021,Gupta2021-2,Pellegrino2016,Levitov2016,Shytov2018,Nazaryan2024,Falkovich2017,Palm2024,AharonSteinberg2022} — the quintessential signature of viscosity.

Several decades passed before electron viscosity made its first — indirect — appearance in experiments \cite{deJong1995}. As Gurzhi envisioned, signatures of viscosity were consistently sought in the relationship between net current and induced voltages — including negative resistance as a possible manifestation of vorticity \cite{Bandurin2016,Gupta2021,Pellegrino2016,Levitov2016,Shytov2018,Nazaryan2024,Falkovich2017}. Yet it was clear from the outset that such approaches would remain inherently indirect and challenging to interpret \cite{Gurzhi1995,Estradalvarez2025,Shytov2018,Polini2020}. First, the simple ohmic relation between current density and electric field no longer holds in the hydrodynamic regime. Second, electron collisions conserve total momentum and therefore do not directly affect the net current. Finally, identifying hydrodynamic signatures is only half the battle: one must also prove their origin is indeed hydrodynamic. Vortices, for example, are also expected in the ballistic regime, but with a fundamentally different origin: geometry, not interparticle scattering \cite{Gupta2021,Berggren1992,Exner1998}. Detecting viscous effects — and vortices in particular — has proven extremely difficult, and claims of their observation in transport measurements are persistently questioned. At the same time, without clear experimental evidence of vortex existence, the hydrodynamic picture of electron transport was not convincing. The need for alternative detection schemes became increasingly clear.

In this work, we offer such an alternative: direct mechanical detection of electron vortices. As circulating flows of charge, vortices carry a magnetic moment. Like a compass needle placed in an external field, they experience a torque. Our approach measures this torque directly — a fundamental shift from conventional resistance measurements.

Among other alternative schemes, recent breakthrough studies using scanning probe techniques \cite{Ku2020,Vool2021} stand out. Using SQUID and diamond NV-center magnetometry, vortices have been directly imaged in graphene \cite{Palm2024} and — unexpectedly — in WTe$_2$ under conditions extending beyond the hydrodynamic and ballistic regimes \cite{AharonSteinberg2022} (the so-called para-hydrodynamic regime). Magnetometry is powerful and direct, yet it has not become a mainstream tool in vortex studies — likely because of its complexity and equipment demands. Returning to the compass analogy: instead of detecting the needle's own field, we apply an external field and measure its motion — that motion is our signal.

To implement this idea, the vortex must be embedded within a mechanical structure small enough to respond to the torque - a thin nanomechanical resonator in our case. This shift — from field detection to force detection — allows us to integrate the electron system and the vortex detector on the same chip, entirely circumventing the complexity of scanning probe techniques.

Our platform — AlGaAs/GaAs heterostructures  — is uniquely suited to this task. These heterostructures are a well-established testbed for quantum transport \cite{Ensslin2025}; inspired by advances in graphene hydrodynamics, researchers have returned to them in search of viscous effects \cite{deJong1995,Keser2021,Gusev2020,Sarypov2025-2,Gupta2021,Gupta2021-2,Alekseev2016}. Crucially, they have also been extensively developed as a platform for micro- and nanomechanical resonators \cite{Yamaguchi2017,Pogosov2022}, providing a well-controlled interface between electronic and mechanical degrees of freedom. In nanoelectromechanical systems, this interface makes it possible to study transport phenomena using mechanical methods \cite{Okazaki2016,Harris2000}, including the processes invisible in transport, such as transfer between coupled quantum dots \cite{Khivrich2019}.

We apply an alternating current close to the mechanical resonant frequency. The oscillating vortex generates a magnetic moment, which drives vibrations in an in-plane magnetic field; the phase of the response reveals the direction of circulation. The technique is sensitive to both hydrodynamic and ballistic vortices, while the temperature evolution of the signal makes it possible to distinguish between the two regimes.

Crucially, this approach does more than just reveal the presence — or nature — of a vortex. Our observations point to a deeper implication: electron viscosity is not merely a transport phenomenon we probe with mechanics — it is the very mechanism shaping the response of nanoelectromechanical systems. We show that the hydrodynamic description of electron flow extends naturally to nanoelectromechanics. In doing so, the vibrant fields of electron hydrodynamics and nanomechanics become inextricably linked.

\subsection*{Design and operation principles of the devices}\label{sec1}

\begin{figure} 
	\centering
	\includegraphics[width=1\textwidth]{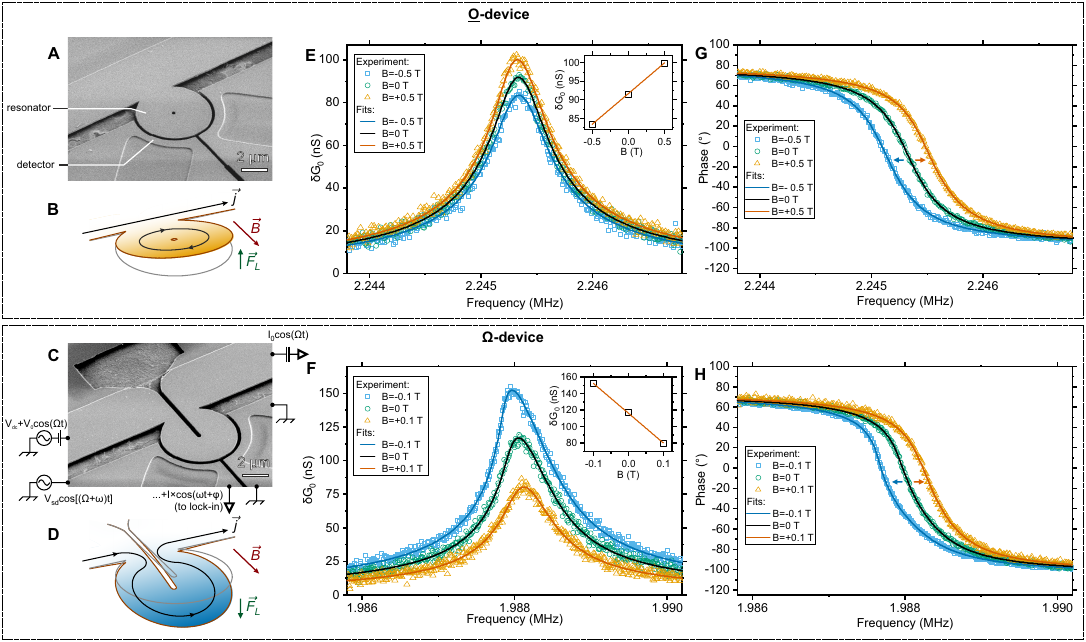} 

	\caption{\textbf{Comparative experiment confirming the vortex presence}
		(\textbf{A}), Scanning electron microscope image of a studied device ($\underline{\mathrm{O}}$-device) - a nanomechanical cantilever-like resonator hosting a circular electron cavity attached to a straight conductive channel. (\textbf{B}), A current flowing through the channel can induce an electron vortex in the cavity accompanied by a counterflow at the free edge. In an in-plane magnetic field $\vec{B}$, Lorentz force $\vec{F}_\mathrm{L}$ contributes to driving of mechanical oscillations (the calculated shape of the first mode is displayed, color shows the displacement). (\textbf{C}), Image of a reference sample ($\Omega$-device), where the vortex formation is geometrically suppressed by an additional etched trench. The common measurement scheme is shown; mechanical oscillations are detected by measuring the conductance response $\delta G$ of a constriction in a two-dimensional electron gas (detector) using heterodyne down-mixing. (\textbf{D}), Co-flow at the free edge leads to a Lorentz force opposite to that in $\underline{\mathrm{O}}$-device sample. (\textbf{E-H}), Amplitude $\delta G_0$ and phase of the measured response as functions of the driving frequency $\Omega/2\pi$ at various magnetic fields. Phase curves are horizontally offset by 200 Hz for clarity. Top panels: $\underline{\mathrm{O}}$-device; bottom panels: $\Omega$-device. The magnetic field leads to opposite effects in samples supporting and preventing vortex formation. Insets in panels \textbf{E,F} show the resonant amplitude as a function of magnetic field.}
	\label{fig:Fig1}
\end{figure}

Experimental samples are fabricated from a GaAs/(Al,Ga)As heterostructure hosting a high-mobility two-dimensional electron gas (2DEG; see Methods \cite{methods}). Current $I$ flowing through a straight 5 $\mu$m-wide channel is used to drive an electron vortex in an adjacent circular cavity with a diameter of 6 $\mu$m (see Fig. \ref{fig:Fig1} A); the system is thus qualitatively similar to those studied in Refs. \cite{AharonSteinberg2022,Palm2024}. A conceptual feature of our system is that the layer beneath the cavity (sacrificial layer) is selectively removed, the 166 nm-thick cavity is freely suspended, and, apart from being electrically functional, it serves as a nanomechanical resonator with cantilever-type geometry (one clamped edge, one free edge).

To reveal the vortex formation, we detect its distinct signature: the electron counter-flow at the free edge of the cantilever (Fig.~\ref{fig:Fig1} B). In an in-plane magnetic field $\vec{B}$, the net current $I$ generates a Lorentz force
\begin{equation}\label{eq:LF}
F_\mathrm{L}= \Lambda IB,
\end{equation}
that bends the cantilever \cite{Schmid2023,Bachtold2022,Cleland1998}.

Here, the prefactor $\Lambda$ is the key experimental observable revealing vorticity. As detailed below, it encodes the spatial distribution of the current density $\vec j(\vec{r})/I$, and its sign is determined by the current direction at the free edge. The fundamental flexural mode of the cantilever acts as a spatial filter: it is maximally sensitive to forces near the free end and minimally sensitive near the clamped edge \cite{Schmid2023}. Thus, the measured $\Lambda$ is dominated by the flow at the free edge---where the counter-flow is expected---while the current in the main channel contributes negligibly. The sign of $\Lambda$ therefore provides an unambiguous, direct mechanical signature of the vortex.

To validate our detection scheme, we compare the described sample which can host a vortex ($\underline{\mathrm{O}}$-device) to a reference sample where vortex formation is geometrically suppressed  ($\Omega$-device), guaranteeing co-flow at the free edge (see Fig. \ref{fig:Fig1} C,D). In the $\Omega$-device, an etched trench extending to the cavity center cuts the main channel and directs the entire electron flow via the cantilever free edge, preventing any counter-flow. Consequently, while a sufficiently strong vortex in the $\underline{\mathrm{O}}$-device yields a Lorentz force with a characteristic sign of $\Lambda$, the reference sample produces a force of the opposite sign. This comparative design provides a direct, built-in control: opposite force signs in the two devices signal that our measurement indeed detects the presence of a vortex in the $\underline{\mathrm{O}}$-sample.

To excite the resonator, we apply voltage $V_\mathrm{dc}+V_0\cos(\Omega t)$ between the channel terminals ($V_\mathrm{dc}=10$ V, $V_0=2.5$ mV), resulting in current $I = I_0 \cos(\Omega t)$ (the dc component is blocked by a 100 nF capacitance). The arising force
\begin{equation}
F=F_0\cos(\Omega t)
\end{equation}
drives the cantilever into flexural oscillation, with the displacement being
\begin{equation}\label{eq:displacement}
u = u_0\cos(\Omega t + \varphi).
\end{equation}
Here $u_0$ and $\varphi$ depend on $\Omega$ in the way typical for a driven harmonic oscillator.

This motion is detected electrostatically \cite{Shevyrin2025,Schmid2023,Bachtold2022,Imboden2014}: the cantilever itself serves as a gate for a nearby 2DEG constriction, biased by $V_\mathrm{dc}$. As the cantilever oscillates, the gate--constriction capacitance varies, modulating the electron density in the constriction and its conductance:
\begin{gather}
G=G_0+\delta G_0\cos(\Omega t+\varphi),\label{eq:G}\\
\delta G_0=\frac{\partial G}{\partial u}u_0.\label{eq:dG0}
\end{gather}
A heterodyne down-mixing scheme \cite{Pogosov2022,Sazonova2004} (see Fig.~\ref{fig:Fig1} C and Methods\cite{methods}) extracts both $\delta G_0$ and $\varphi$ (up to a constant instrumental phase shift), providing a direct readout of the mechanical response.

\subsection*{Distinguishing co-flow and counter-flow mechanically}\label{sec2}

Fig. \ref{fig:Fig1} E-H shows the measured resonant response: $\delta G_0$ follows the Lorentzian lineshape of the mechanical mode (see Eq. \ref{eq:dG0}):
\begin{equation}
\delta G_0=\frac{\partial G}{\partial u}\frac{F_0Q}{m\Omega_0^2}\times\frac{1}{\sqrt{1+\left[2Q(\Omega-\Omega_0)/\Omega_0\right]^2}},
\end{equation}
with weak nonlinearity \cite{Schmid2023,Bachtold2022} at high amplitudes. Here $Q$ is the quality factor, $m$ is the effective modal mass, $\Omega_0$ is the resonant frequency. The results are shown for both $\underline{\mathrm{O}}$- and $\Omega$-devices, measured at $T=4$ K. Most importantly, the magnetic field effects are opposite in these two samples: while $\partial(\delta G_0)/\partial B>0$ in $\underline{\mathrm{O}}$-device, in $\Omega$-device $\partial(\delta G_0)/\partial B<0$. This contrast is the expected signature of counter-flow versus co-flow. Since co-flow is guaranteed in the $\Omega$-device, the opposite sign in the $\underline{\mathrm{O}}$-device directly confirms the presence of a vortex.

The signal is noticeably non-zero at $B=0$, which means that, apart from the Lorentz force, there is an additional physical mechanism of the oscillation driving. More detailed experiments described in Methods \cite{methods} show that this mechanism is electrostatic in nature \cite{Shevyrin2025,Shevyrin2021}, arising from the capacitive coupling between the cantilever and the surrounding electrodes. The total force is
\begin{equation}
F=(F_\mathrm{L0}+F_\mathrm{ES0})\cos(\Omega t),
\end{equation}
where
\begin{gather}
F_\mathrm{ES0}=\frac{\partial C}{\partial u}\frac{V_0}{2}(V_\mathrm{dc}-V_1)\label{eq:ESFA},\\
F_\mathrm{L0}=\Lambda I_0B\label{eq:LFA}
\end{gather}
are the amplitudes of the electrostatic and Lorentz forces, $C$ is the capacitance between the cavity and surroundings, and $V_1$ is the charge neutrality voltage caused by built-in charge \cite{Shevyrin2021} (equal to $-1.8$ and $-1.4$ V for $\underline{\mathrm{O}}$- and $\Omega$-devices, respectively).

The opposite signs of $\partial(\delta G_0)/\partial B$ at $T = 4$ K (Fig.~\ref{fig:Fig1} E-H) already indicate the presence of a vortex in the $\underline{\mathrm{O}}$-device. To conclusively rule out alternative explanations, we subject this conclusion to a further test: if the signal indeed originates from a vortex, its sign must reverse at elevated temperatures, converging to the sign of the co-flow-dominated $\Omega$-device.

The reasoning is straightforward. In the Ohmic regime at high temperatures, vortices cannot exist---any counter-flow in the $\underline{\mathrm{O}}$-device should vanish, turning into co-flow. The $\Omega$-device, where co-flow is geometrically enforced, retains it at all temperatures. Thus, if our interpretation is correct, the two devices should exhibit opposite Lorentz force signs at low $T$ but the \emph{same} sign at high $T$, when the vortex is suppressed. The key quantity governing this crossover is $\Lambda(T)$, which encodes the current distribution at the cantilever edge. Extracting $\Lambda(T)$ from the raw data $\delta G_0(B,T)$ requires separating several temperature-dependent factors---a step we now describe.

The resonant amplitude $\delta G_0^*$, which we obtain by fitting $\delta G_0(\Omega)$, is
\begin{equation}\label{eq:dG0*}
\delta G_0^*=\frac{\partial G}{\partial u}\frac{Q}{m\Omega_0^2}(F_\mathrm{L0}+F_\mathrm{ES0}).
\end{equation}
It contains two temperature-dependent prefactors: the detector responsivity $\partial G/\partial u$ and the quality factor $Q$ (here $F_\mathrm{L0}$ and $F_\mathrm{ES0}$ are the amplitudes of the Lorentz and electrostatic forces, respectively). The quality factor \cite{Schmid2023,Imboden2014} is readily obtained from the width of the resonance curves and poses no difficulty. The responsivity $\partial G/\partial u$, however, is unknown without calibration — and it is this factor that prevents a direct extraction of $\Lambda(T)$ from the raw data.

To eliminate the unknown $\partial G/\partial u$, we use the electrostatic force as a convenient reference: it is independent of both temperature and magnetic field. The relative change in the measured signal induced by magnetic field directly gives the ratio of magnetomotive and electrostatic forces:
\begin{equation}
\frac{\delta G_0^*(B)-\delta G_0^*|_{B=0}}{\delta G_0^*|_{B=0}}=\frac{F_\mathrm{L0}}{F_\mathrm{ES0}}.
\end{equation}
We take $B$-derivative of this equation and, as $F_\mathrm{L0}\propto I_0$ and $F_\mathrm{ES0}\propto V_0$, do further normalization to obtain the following quantity from the experiment:
\begin{equation}
\Gamma=\frac{V_0}{2I_0}\frac{1}{\delta G_0^*|_\mathrm{B=0}}\frac{\partial (\delta G_0^*)}{\partial B}.
\end{equation}
Up to a constant temperature-independent factor, $\Gamma(T)$ directly reflects the temperature dependence of $\Lambda$ (see Eqs.~\ref{eq:ESFA} and~\ref{eq:LFA}) — the quantity encoding the vortex circulation:
\begin{equation}
\Gamma(T)=\left[\frac{\partial C}{\partial u}(V_\mathrm{dc}-V_1)\right]^{-1}\Lambda(T).
\end{equation}

The measured $\Gamma(T)$ dependences are shown in Fig. \ref{fig:Fig2} for $\underline{\mathrm{O}}$- and $\Omega$-devices. The $\Omega$-device exhibits only weak temperature dependence, and the sign of $\Gamma$ remains unchanged — consistent with co-flow at the cantilever edge at all temperatures. In $\underline{\mathrm{O}}$-device, $\Gamma(T)$ varies strongly and changes sign near $19$ K, matching the sign of $\Omega$-device at higher temperatures. Qualitatively, this is exactly the behavior expected from a vortex suppression by temperature. In the next section, we consider it quantitatively.

\subsection*{Vortices: ballistic or hydrodynamic?}\label{sec3}

In this section we first establish the link between the phenomenological key parameter $\Lambda$ and the microscopic electromechanical description, applicable in a general case. Next, we calculate $\Lambda(T)$ in the hydrodynamic mode and, finally, compare the theoretical and experimental results to reveal the difference caused by ballistic effects.

The cantilever's flexural vibration is described by a displacement field $U(\vec{r},t)$ factorized as
\begin{equation}
U(\vec{r},t)=u(t)\psi(\vec{r}),
\end{equation}
where $\psi(\vec{r})$ is the dimensionless mode shape normalized to the maximum deflection. The net effective driving force for this mode is the overlap integral \cite{Schmid2023} of the Lorentz force density with the mode shape:
\begin{equation}
\vec{F}_\mathrm{L}(t)=\int\left[\vec{j}(\vec{r},t)\times \vec B\right]\psi(\vec{r})\ \mathrm{d^2}\vec{r}.
\end{equation}
In the linear approximation, the current density $\vec{j}$ is also factorized. Its component orthogonal to the magnetic field is
\begin{equation}
j_x(\vec{r},t)=I(t)\chi(\vec{r}).
\end{equation}
Thus, the out-of-plane projection of $\vec{F_\mathrm{L}}$ is
\begin{equation}
F_\mathrm{L}=\left(\int\psi(\vec{r})\chi(\vec{r})\mathrm{d}^2\vec{r}\right) IB.
\end{equation}
Comparison with Eq. \ref{eq:LF} makes it obvious that
\begin{equation}
\Lambda(T)=\int \psi(\vec r)\chi\left(\vec r,T\right)\mathrm{d^2}\vec{r}.
\end{equation}

\begin{figure}
\centering
\includegraphics[width=0.6\textwidth]{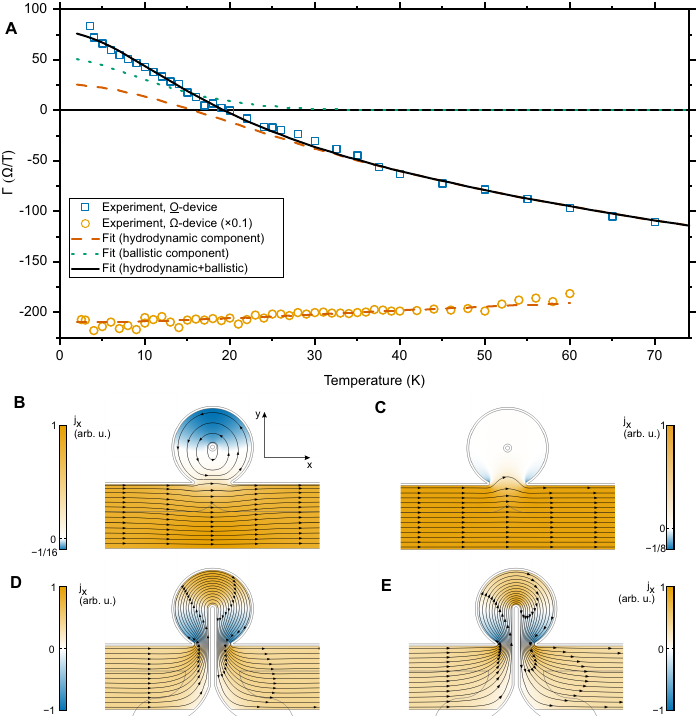}
\caption{\label{fig:Fig2} \textbf{Temperature evolution of vortices } (\textbf{A}) Temperature dependence of coefficient $\Gamma\propto\Lambda$ characterizing the influence of the spatial current distribution on the Lorentz force $F_\mathrm{L}=\Lambda BI$. Hydrodynamic fits based on Stokes equation solution are shown. In $\underline{\mathrm{O}}$-device, additional ballistic component is needed to fit the data at low temperatures. (\textbf{B-E}), Calculated current distributions corresponding to low (left panels) and high (right panels) temperatures. The corresponding points are labeled in panel (\textbf{A}).}
\end{figure}

We calculate $\psi(\vec{r})$ in the thin plate approximation using the finite element method, see Fig. \ref{fig:Fig1} B,D. To calculate $\chi(\vec r,T)$ in the hydrodynamic and ohmic regimes (see Fig. \ref{fig:Fig2} B-E), we use a common approach \cite{Polini2020,AharonSteinberg2022} and self-consistently solve the linearized Stokes equation and the continuity equation (see Methods \cite{methods}):
\begin{eqnarray}
\vec{j}=-\sigma\vec\nabla\phi+D^2\Delta \vec{j},\label{eq:j}\\
\mathrm{div}\vec{j}=0.\label{eq:cont}
\end{eqnarray}
where $D=\sqrt{l_\mathrm{ee}l}/2$ is the Gurzhi length which quantifies viscous effects, $l$ is the transport mean free path (measured), $l_\mathrm{ee}$ is the electron-electron scattering mean free path (evaluated using the commonly used Giuliani-Quinn model \cite{Giuliani1984}), $\sigma$ is conductivity, and $\phi$ is electric potential.

Comparison of the measured $\Gamma(T)$ and simulated $\Lambda(T)$ dependences shows that, for $\underline{\mathrm{O}}$-device, the hydrodynamic model describes the experiment reasonably well at high temperatures, but fails to do so at low temperatures, consistently with the breakdown of hydrodynamics when $l_\mathrm{ee}$ exceeds other length scales. Note that the experimentally measured temperature where $\Gamma(T)=0$ (19 K) is in a reasonable agreement with the theoretical value $T=$16 K. In $\Omega$-device, the hydrodynamic model predicts weak temperature dependence, similar to that observed experimentally across the entire temperature range.

The systematic deviation of the hydrodynamic model from the experimental $\Gamma(T)$ at low temperatures (Fig.~\ref{fig:Fig2} A) suggests the emergence of an additional contribution. A natural candidate is ballistic transport: at low temperatures, long mean free paths allow electrons to traverse the cavity along isolated trajectories, forming ballistic vortices distinct from their hydrodynamic counterparts. A full treatment of the ballistic-to-hydrodynamic crossover would require solving the Boltzmann equation \cite{Gupta2021,Egorov2025} — a complex task beyond the scope of this work. Instead, we adopt a minimal phenomenological description that captures the essential physics.

We assume that the ballistic contribution arises from electrons traveling along isolated paths. Increasing temperature enhances scattering \cite{Gupta2021-2,Egorov2025}, progressively suppressing such ballistic transport. The survival probability for an electron to complete a path of characteristic length $L$ scales as $\exp(-L/l_\mathrm{scat})$, where $l_\mathrm{scat}^{-1}=l_\mathrm{ee}^{-1}+l^{-1}$. Accordingly, we model the total response as a sum of hydrodynamic and ballistic terms:
\begin{equation}
\Gamma(T) = c_1 \cdot \Lambda(T) + c_2 \cdot \exp\left(-\frac{L}{l_\mathrm{scat}(T)}\right),
\end{equation}
with $c_1$, $c_2$, and $L$ as fitting parameters. The best fit yields $L \approx 4.5\ \mu\mathrm{m}$, comparable to the cavity diameter and consistent with ballistic trajectories spanning the structure. The ballistic term dominates below $20$ K but becomes negligible above $30$ K, where $\Gamma(T)$ follows the purely hydrodynamic behavior. This agreement supports the coexistence of two distinct vortex types and demonstrates that nanomechanical detection, combined with temperature-dependent measurements, can discriminate between them. The strong temperature dependence of $\Lambda(T)$ — and hence of the driving force — implies that the mechanical response itself is shaped by electron viscosity. In this sense, the resonator does not merely probe the electron fluid; it is actuated by it.

\subsection*{Conclusions and outlook}

We have demonstrated that a nanomechanical resonator can directly detect electron vortices via the torque exerted on the circulating current in a magnetic field. A comparative experiment with geometrically suppressed vortices provides unambiguous identification, while temperature-dependent measurements reveal a smooth crossover from ballistic to hydrodynamic behavior in a GaAs-based two-dimensional electron gas. The observed vortices, with diameters of several micrometers, exceed in size those previously reported in graphene \cite{Palm2024}, highlighting the material-specific scales of electron hydrodynamics.

These results also establish a new perspective: while viscous electron effects are notoriously difficult to detect in transport measurements, their impact on nanoelectromechanical systems can be substantial and play a previously unrecognized role. Graphene, an already developed platform for both hydrodynamic studies and nanomechanical resonators, offers another natural testing ground, where at submicron scales, viscous effects may persist even at room temperature.

Beyond hydrodynamics, ballistic vortices are of interest in their own right. Early work predicted behavior reminiscent of turbulence in quantum nanostructures — a striking possibility given the low temperatures involved \cite{Berggren1992}. The magnetic moment associated with quantum vortices, identified theoretically \cite{Exner1998}, now becomes accessible to direct measurement via nanomechanical torque detection.

Implementing our sensing scheme in other two-dimensional materials could open new avenues for exploring electron hydrodynamics and quantum transport alike. Dispersive measurements \cite{LaHaye2009,Khivrich2019} based on resonant frequency shifts may offer higher precision than the amplitude detection used here. At gigahertz-range or higher frequencies, where the mechanical period approaches the characteristic timescales of vortex rotation and decay, strong coupling between mechanical motion and electron vortices is expected, enabling direct access to vortex kinetics. Looking further, nanomechanical detection may provide a window into more complex collective states, such as the long-predicted onset of genuine turbulence or pre-turbulence in an electron fluid \cite{Mendoza2011,Mller2009}.




\clearpage
\bibliography{Shevyrin_vortices}
\bibliographystyle{sciencemag}

\section*{Acknowledgments}

The authors thank Maxim S. Aksenov, Natalya R. Vicina, Sergey V. Ishutkin and Ekaterina V. Anishenko - for providing the access to fabrication facilities - and Evgeniy Yu. Zhdanov - for helping with cryogenic measurements. Scanning electron microscope images are obtained at the Center for Collective Use "Nanostructures". 

\paragraph*{Funding:}
The work is supported by Russian Science Foundation (grant~№22-12-00343-$\Pi$). Transport characterization of the heterostructures is supported by the state assignment of Ministry of Science and Higher Education of the Russian Federation (grant~№FWGW-2025-0023).

\newpage

\renewcommand{\thefigure}{S\arabic{figure}}
\renewcommand{\thetable}{S\arabic{table}}
\renewcommand{\theequation}{S\arabic{equation}}
\renewcommand{\thepage}{S\arabic{page}}
\setcounter{figure}{0}
\setcounter{table}{0}
\setcounter{equation}{0}
\setcounter{page}{1}

\begin{center}
\section*{Supplementary Materials for\\ \scititle}

Andrey~A.~Shevyrin$^{1\ast}$,
	Askhat~K.~Bakarov$^{1}$,
	Arthur~G.~Pogosov$^{1,2}$\\
	\small$^{1}$Rzhanov Institute of Semiconductor Physics, Novosibirsk, 630090, Russia.\\
	\small$^{2}$Department of Physics, Novosibirsk State University, Novosibirsk, 630090, Russia.\\
	\small$^\ast$Corresponding author. Email: shevandrey@isp.nsc.ru
\end{center}

\newpage

\subsection*{Materials and Methods}

\subsubsection*{Samples fabrication}
The heterostructure \cite{Pogosov2022,Shevyrin2025} is grown using molecular beam epitaxty on a GaAs (100) semi-insulating wafer. On top of the heterostructure, the following layers reside (in the order of growth): Al\textsubscript{0.8}Ga\textsubscript{0.2}As sacrificial layer (400 nm), AlAs/GaAs superlattice (42.5 nm, alternating 4 AlAs/8 GaAs monolayers), Si $\delta$-doping ($1.5\times 10^{11}$ cm\textsuperscript{-2}), AlAs/GaAs superlattice (33.5 nm), GaAs quantum well (13 nm), AlAs/GaAs superlattice (33.5 nm), Si $\delta$-doping ($1.5\times 10^{11}$ cm\textsuperscript{-2}), AlAs/GaAs superlattice (33.5 nm), GaAs (5 nm), Si $\delta$-doping ($2.5\times 10^{11}$ cm\textsuperscript{-2}), GaAs (5 nm). The layers above the sacrifical layer form a 166 nm-thick membrane with a two-dimensional electron gas in the middle of its thickness.

Shallow mesa and Ohmic contacts are patterned using mask-free optical lithography (direct writer). The mesa is wet etched in H\textsubscript{2}SO\textsubscript{4}:H\textsubscript{2}O\textsubscript{2}:H\textsubscript{2}O 1:8:80 solution to the depth of 80 nm. Among others, the shallow mesa is used to form the oscillation detectors. The Ohmic contacts consist of Au/Ge eutectic alloy, with Ni adhesion layer. The geometry of the resonators is defined using electron beam lithography with subsequent anisotropic etching in BCl\textsubscript{3}:Ar plasma to the depth of approximately 180 nm. After that, the sacrificial layer is selectively removed from beneath the resonators in HF:H\textsubscript{2}O 1:10 solution. After rinsing in deionized water, the samples are dried in critical carbon dioxide to avoid capillary forces.

\subsubsection*{Measurement setup}

The measurements are performed in a dry He-4 cryostat with a variable temperature insert equipped with a superconducting magnet. During the measurements, the samples are placed in vacuum to avoid damping of mechanical oscillations due to environment \cite{Imboden2014}.

Each device is equipped with two constrictions-detectors. Only one of them is used, another one is fully grounded. Mechanical oscillations modulate conductance of the constriction (see Eqs. \ref{eq:displacement}, \ref{eq:G}, \ref{eq:dG0}). To probe the response amplitude $\delta G_0$ and phase $\varphi$, we apply an ac voltage $\delta V_\mathrm{sd}=\delta V_\mathrm{sd0}\cos[(\Omega+\omega)t]$ (where $\omega/2\pi=9.14$ kHz$\ll\Omega/2\pi$, $\delta V_\mathrm{sd0}=28$ mV, linear regime) to one of the terminals of the constriction. Another terminal is connected to an input of a transimpedance preamplifier (virtual ground) which, in turn, is connected to an input of a lock-in amplifier. The current flowing through the constriction is $I=(G_0+\delta G)\delta V_\mathrm{sd}=\ldots+\delta I_\omega\cos(\omega t-\varphi)$; the constriction plays the role of an on-chip heterodyne down-mixer \cite{Sazonova2004,Shevyrin2025}. The amplitude of the current measured at low frequency $\omega$ is proportional to $\delta G_0$, its phase is the opposite to the oscillation phase up to a constant instrumental shift. The response amplitude is extracted as $\delta G_0=2\delta I_\omega/\delta V_\mathrm{sd0}$. The current $I_0$ flowing through the main channel is also measured using the second input of the transimpedance amplifier.

\subsubsection*{Driving and detection}

In this section, we first describe a simplified picture of the factors influencing the measured signal $\delta G$ (see Eq. \ref{eq:dG0*}) and then show experimental data - dependences on $V_\mathrm{dc}$ and $B$ - confirming this description.

To detect mechanical oscillations, we use the electrostatic method \cite{Sazonova2004,Schmid2023,Bachtold2022}. The resonator serves as a gate for the constriction-detector; the dc component of the gate voltage is $V_\mathrm{dc}$. Due to the system symmetry, the amplitude of the ac potential of the cavity is $V_0/2$ (the amplitude of the measured signal remains unchanged when the contacts are swapped at $B=0$). The resonator displacement $u$ changes the resonator-constriction capacitance, thus modulating electron density in the detector and its conductance. In the simplest approximation, the constriction responsivity should linearly depend on $V_\mathrm{dc}$. Due to the presence of a built-in charge (donors, surfaces, depletion regions, piezoelectric charge), the constriction responsivity is non-zero at $V_\mathrm{dc}=0$; voltage $V_2$ compensates this effect:
\begin{equation}
\frac{\partial G}{\partial u}=\alpha(V_\mathrm{dc}-V_2).
\end{equation}

\begin{figure} 
	\centering
	\includegraphics[width=1\textwidth]{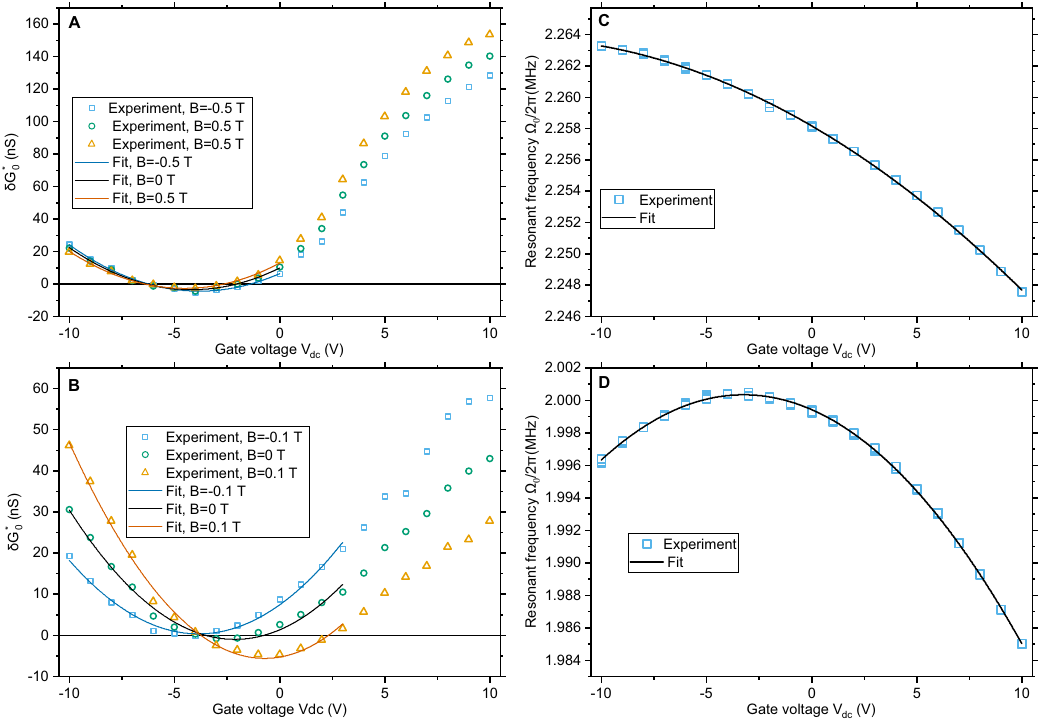} 

	\caption{\textbf{Gate voltage dependence of the measured signal.} Amplitudes (\textbf{A,B}) and resonant frequencies (\textbf{C,D}) measured as functions of the dc component of the gate voltage. Negative amplitude values correspond to a 180$^\circ$ phase shift. Top and bottom panels correspond to $\underline{\mathrm{O}}$- and $\Omega$-devices, respectively.}
    \label{figext:figext1}
\end{figure}

Substituting this into Eq. \ref{eq:dG0*}, one can expect parabolic dependence of the resonant amplitude $\delta G_0^*$ on $V_\mathrm{dc}$:

\begin{equation}\label{eq:dG0*e}
\delta G_0^*=\frac{\alpha QV_0}{m\Omega_0^2}\left[\frac{1}{2}\frac{\partial C}{\partial u}(V_\mathrm{dc}-V_1)+\Lambda BG_\mathrm{ch}\right]\times\left[V_\mathrm{dc}-V_2\right],
\end{equation}
where $G_\mathrm{ch}=I_0/V_0$ is the main channel conductance. The corresponding $\delta G_0^*(V_\mathrm{dc})$ curves measured at various $B$ values agree with this expected dependence (see Fig. \ref{figext:figext1}). For both $\underline{\mathrm{O}}$- and $\Omega$-devices, there are voltages $V_2=-2.3$ V, $-6.3$ V where the response is zero and the curves measured at different $B$ values intersect (the right square bracket is zero in Eq. \ref{eq:dG0*e}). These voltages correspond to zero responsivity of the detector. For each curve, another voltage at which $\delta G_0^*=0$ corresponds to zero force (left square brackets in Eq. \ref{eq:dG0*e}). This voltage depends linearly on magnetic field. The curves measured at various $B$ values can be fitted by parabolas with the same quadratic coefficients. This means that the value $\alpha Q/m\Omega_0^2$ is independent of magnetic field. At large positive $V_\mathrm{dc}$ values the experimental data deviate from parabolas.

The curves in Fig. \ref{figext:figext1} C,D show that the measured resonant frequencies $\Omega_0/2\pi$ depend on $V_\mathrm{dc}$ in a perfectly parabolic way. This is the manifestation of the electrostatic spring softening effect \cite{Bachtold2022}, typical for resonators electrostatically interacting with surrounding electrodes:
\begin{equation}
\Delta{\Omega_0}=-\frac{\partial^2C}{\partial u^2}\frac{(V_\mathrm{dc}-V_3)^2}{4m\Omega_0}.
\end{equation}
The fact that $\Omega_0(V_\mathrm{dc})$ curves are perfectly fitted in the entire gate voltage range means that capacitance $C$ is independent of the gate voltage. Consequently, the deviation from parabolas observed in $\delta G_0^*(V_\mathrm{dc})$ dependences at large positive gate voltage values is most likely related to nonlinearity of the responsivity $\partial G/\partial u$ as a function of $V_\mathrm{dc}$. We note that this does not affect the results presented in the manuscript, since the detection is eliminated when we consider $\Gamma(T)$.

\subsubsection*{Transport parameters measurement}

\begin{figure} 
	\centering
	\includegraphics[width=1\textwidth]{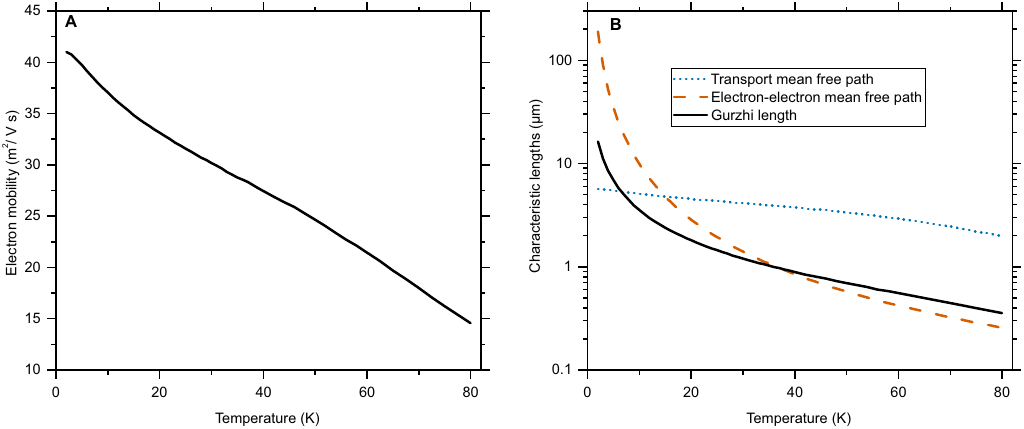} 

	\caption{\textbf{Transport parameters.} (\textbf{A}), Electron mobility measured in 50$\times$20 $\mu$m suspended Hall bars. (\textbf{B}) Characteristic transport length scales extracted from the measurements.}
    \label{figext:figext2}
\end{figure}

Electron density $n\approx6.8\times10^{11}$ cm$^{-2}$ and mobility $\mu$ are extracted as functions of temperature from the dependences of longitudinal and Hall resistances \cite{Zhdanov2026} on a perpendicular magnetic field measured in 50$\times$20 $\mu$m suspended Hall bars made from the same heterostructure. The density values extracted from the Hall resistance and Shubnikov-de-Haas oscillations coincide. The results are presented in Fig. \ref{figext:figext2}.

The transport mean free path calculated as $l=\hbar\mu\sqrt{2\pi n}/e$, and the electron-electron mean free path calculated using the Giuliani-Quinn model \cite{Giuliani1984}
\begin{equation}
l_\mathrm{ee}^{-1}=\frac{(k_\mathrm{B}T)^2}{hv_\mathrm{F}E_\mathrm{F}}\left[\ln\left(\frac{E_\mathrm{F}}{k_\mathrm{B}T}\right)+\ln\left(\frac{2q_\mathrm{TF}}{k_\mathrm{B}T}\right)+1\right]
\end{equation}
are displayed in Fig. \ref{figext:figext2} B). These dependences are used to calculate the Gurzhi length $D(T)=\sqrt{l_\mathrm{ee}l}/2$.

\subsubsection*{Numerical simulations}

The linearized Stokes equation is solved together with the continuity equation (see Eqs. \ref{eq:j}, \ref{eq:cont}) using finite element method in the same way as described in \cite{AharonSteinberg2022}. The boundary conditions are introduced in terms of slip length $\xi$:
\begin{eqnarray}
\xi\frac{\partial j_\tau}{\partial r_n}=j_\tau,\\
j_n=0,
\end{eqnarray}
where $j_\tau$ and $j_n$ are the components of current density tangential and normal to a boundary, $\partial j_\tau/\partial r_n$ is $j_\tau$ spatial derivative taken along the normal.
The two parameters used in calculations are $D$ and $\xi$; the current density $\vec{j}(\vec{r})$ is normalized by the total current.
The modal shape and eigenfrequencies are calculated using the same mesh using the thin-plate model.
The fitting curves displayed in Fig. \ref{fig:Fig2} A are calculated using the no-stress condition \cite{Sarypov2025-2} ($\xi\rightarrow\infty$). The calculated dependences on the Gurzhi length are converted into dependence on temperature using $D(T)$ described in the previous subsection.

\clearpage 

\end{document}